# An "ad-hoc" modified Likelihood Function Applied to Optimization of Data Analysis in Atomic Spectroscopy.

*Leonardo Bennun

*Applied Physics Laboratory, Physics Department, University of Concepcion; Casilla 160-C, Concepcion, Chile. E-mail: lbennun@udec.cl (L. Bennun)

**Abstract:** In this paper we propose an "*ad—hoc*" construction of the Likelihood Function, in order to develop a data analysis procedure, to be applied in atomic and nuclear spectral analysis.

The classical Likelihood Function was modified taking into account the underlying statistics of the phenomena studied, by the inspection of the residues of the fitting, which should behave with specific statistical properties. This new formulation was analytically developed, but the sought parameter should be evaluated numerically, since it cannot be obtained as a function of each one of the independent variables. For this simple numerical evaluation, along with the acquired data, we also should process many sets of external data, with specific properties — This new data should be uncorrelated with the acquired signal.

The developed statistical method was evaluated using computer simulated spectra. The numerical estimations of the calculated parameter applying this method, indicate an improvement over accuracy and precision, being one order of magnitude better than those produced by least squares approaches.

We still have to evaluate the improvement produced by this method over Detection and Quantitation Limits, in TXRF spectral analysis.

**INTRODUCTION**

In atomic and nuclear spectroscopies, almost all of the methods for spectral analysis are based on the least squares algorithms. These data processing and interpretation techniques are usually applied indistinctly to linear or non—linear physical systems. The great success of the application of this method is based essentially on a deep agreement between the underlying physical phenomena studied and the applied mathematical theory.

In most of the atomic and nuclear spectroscopic techniques, in order to obtain the sample/system characterization, we found a common sequence of events, which can be detailed as follows: *a*) a generation of energetic particles required as an excitation source, *b*) the de-excitation process of the sample, *c*) the detection (usually trough a solid state detector) and *d*) an acquisition process (typically the electronic chains are composed by a preamplifier, amplifier, and a multichannel analyzer). The intrinsic fluctuations of the excitation and de-excitation of the sample, being discrete events, are ruled by the Poisson´s Distribution like all of the atomic or nuclear interactions. Moreover, in the classical texts of Probability, the radioactive decay and nuclear decay reactions are used as iconic examples of the Poisson´s Statistics. This probability also rules the characteristic backgrounds proper of the matrix of the sample. At this stage, we should mention that the Poisson´s Distribution, when it is applied to a large number of events (usually, $n>30$) is very well described by a Normal Distribution. So, the joint probability that describes all kind or sequence of atomic or nuclear events in the sample, is a Normal Distribution. The Monte Carlo methods apply this property in order to infer the average behavior of the particles in the physical system from the average behavior of simulated particles.[1]

At each one of the rest of the steps that follow the electrical signal leaving the detector until it is processed in the multichannel analyzer, it is affected by characteristics fluctuations, like temperature and gain voltage variations, electrical environmental noise, small systematics errors, etc. Again, by applying the Central Limit theorem, we can assure that the Distribution that rule the acquired atomic or nuclear signal, at each channel in the multichannel analyzer, would be a Normal Distribution.

The Central Limit theorem is implicitly applied in the data processing of a large number of complex systems, which are studied with the least squares approach and many of their multiple variations.[2] These methodologies are applied with many strategies in computational mathematics, in order to optimize the results in scientific and engineering applications. Some of the areas of applications are: image and video processing, medical treatments, etc.[3]

In the specific case of spectral analysis, the application of the Maximum Likelihood formulation is almost perfectly suited, since it is constructed considering a Normal Distribution at each channel. But, a question remains: How good is the quality of the results obtained from a Maximum Likelihood estimation for a given parameter? What the Likelihood Function is computing is how likely the measured data is to have come from the distribution assuming a particular value for the hidden parameter; the more likely this is, the closer one would think that this particular choice for hidden parameter is to the true value. So, in atomic spectroscopies, the results obtained from least squares algorithms should be the best, and no method of improvement of the results could be proposed. Moreover, the properties of the obtained results were largely studied. The evaluated parameters from the least squares formulation are unbiased (in the limit of infinite measurements) and have minimum variance among all unbiased linear estimators. This means that the estimates "get us as close to the true unknown parameter values as we can get". For these reasons, an improvement on the quality of the results over those obtained by the least squares method seems to be unrealistic. Moreover, these results are considered as the limit of highest quality, to which tend the results, for instance, obtained from the Neural Network approach [4,5] when they are applied to data analysis in atomic or nuclear spectroscopies, and related systems.

However, in another work [6] we devised a new smoothing method which was applied to simulated spectroscopic data, producing results with better accuracy than those obtained from the least squares approaches.

In this paper, we propose a modification in the construction of the Likelihood function, which leads to a remarkable improvement on the quality of the results provided by the least squares approaches. This modification was made taking into account the underlying statistics of the phenomena studied, by the inspection of the residues of the fitting, which should behave with specific statistical properties. This new formulation was analytically developed, but the calculated parameter should be evaluated numerically, since it cannot be obtained as a function of each one of the independent variables. For the required numerical evaluation, along with the acquired spectrum, we should process many sets of external data with specific properties. This arbitrary term is a random sequence of data which is uncorrelated with the acquired signal. It should be ruled by a Gaussian distribution, having mean value zero and standard deviation Δ=1.

This statistical method was evaluated using computer simulated spectra. The numerical estimations of the calculated parameter applying this method, indicate an improvement over accuracy and precision, one order of magnitude better than those produced by the least squares approaches.

We still have to evaluate the improvement produced by this method over Detection and Quantitation Limits, in TXRF spectral analysis.

**THEORETICAL**

A Maximum Likelihood estimate for some hidden parameter $\gamma$ (or parameters, plural) of some probability distribution is a number $\hat{\gamma}$ computed from an Independent and Identically Distributed sample (IID) $M_1, ..., M_n$ from the given distribution that maximizes something called the "Likelihood Function". Let´s suppose that this distribution is governed by a probability density function (pdf) $G(X; \gamma_1, ..., \gamma_k)$, where the $\gamma_i$'s are all hidden parameters. The Likelihood Function associated to this sample is:

$$L(M_1, ...., M_n) = \prod_{i=1}^{n} G(M_i, \gamma_1, ..., \gamma_k) \quad (1)$$

Note that in all cases the estimated values are represented by English letters while parameter values are represented by Greek letters.

If the distribution is $N(\mu,\sigma^2)$, the Likelihood Function is:

$$L(M_1, ...., M_n; \hat{\mu}, \hat{\sigma}_i^2) = \frac{1}{(2\pi)^{n/2}} \left( \frac{e^{-\frac{1}{2\hat{\sigma}_1^2}(M_1-\hat{\mu}_1)^2}}{\hat{\sigma}_1} \times ... \times \frac{e^{-\frac{1}{2\hat{\sigma}_n^2}(M_n-\hat{\mu}_n)^2}}{\hat{\sigma}_n} \right) \quad (2)$$

where the symbol "^" over the variables $\mu_i$ and $\sigma_i^2$ indicates that they are estimators of their real values.

In atomic spectroscopies (like TXRF, μSR-XRF, PIXE, etc.) the $\hat{\mu}_i$ values could be linearly related with an specific known function, that is, $\mu_i = \alpha F_i$. The function $F$ can be understood as a particular perfectly defined signal, directly linked to the presence of a given element in the sample.[7,8] The $F_i$ sequence take specific values at each channel $i$; and the $\alpha$ value is linearly related mainly with the abundance of this element, and others fundamental parameters.[9] In this case, the Maximum Likelihood yields [10]:

$$lnL = -\sum_{i=1}^{n} \frac{(m_i - \alpha F_i)^2}{2\sigma_i^2} + \sum_{i=1}^{n} ln\left(\frac{1}{\sqrt{2\pi}\sigma_i}\right) \quad (3)$$

At Eq. (3) the $m_i$ values represent the discrete data acquired with the multichannel analyzer; and the $\sigma_i$ values are their standard deviation, at each channel $i$. In this case, the total number of channels analyzed is $n$.

The most probable $\hat{\alpha}$ value with its confidence interval can be calculated from Eq. (3) as [10]:

$$\hat{\alpha} = \frac{\sum_{i=1}^{n} \frac{m_i F_i}{\sigma_i^2}}{\sum_{i=1}^{n} \frac{F_i^2}{\sigma_i^2}} \pm \frac{1}{\sqrt{\sum_{i=1}^{n} \frac{F_i^2}{\sigma_i^2}}} \quad (4)$$

The same result is obtained applying least squares minimization to this heteroscedastic system.

As was discussed in the Introduction, in atomic and nuclear spectroscopies, it is advisable to mention that the results at Eq. (4) are obtained from an adequate representation of the studied physical system. Each one of the underlying processes are indeed well described by the probabilities applied (Normal Distributions) in the developed model. In this sense, the values obtained at Eq. (4) should be the best, and no method of improvement of the results could be proposed. If we ask: Could the quality (accuracy and uncertainty) of a measurement be improved beyond that the Maximum Likelihood Criterion establishes in atomic spectroscopy? Being these spectroscopies well described by the proposed statistical model, the answer should be negative.

But, let´s inspect the residues of the adjustment, which also should behave with specific statistical properties. Once the $\alpha$ parameter is evaluated, the difference $\varepsilon_i = m_i - \alpha F_i$, should on the average: *i*) have equal number/quantity of positive and negative values, *ii*) since it is ruled by a Poisson statistics, its uncertainty at each channel $i$ should be: $\Delta \varepsilon_i = \sqrt{\alpha F_i}$. Taking into account these elements we can propose an improved version of the likelihood function [^T][11,12,13,14] which includes the statistical properties of the studied system, as:

$$lnL = -\sum_{i=1}^{n} \frac{(m_i - \alpha F_i - bg_i\sqrt{\alpha F_i})^2}{\sigma_i^2} + \sum_{i=1}^{n} ln\left(\frac{1}{\sqrt{2\pi}\sigma_i}\right) \quad (5)$$

At Eq. (5) the term $bg_i$ is an arbitrary random sequence of $n$ datum, uncorrelated with the acquired signal. This data is ruled by a Gaussian distribution, having mean value zero and standard deviation $\Delta=1$. This external data is processed simultaneously with the acquired signal.

Now we derivate the Eq. (5) with respect to $\alpha$, in order to obtain the most probable $\alpha$ value, which maximize $L$.

$$\sum_{i=1}^{n}\frac{m_i F_i}{\sigma_i^2} - \alpha\sum_{i=1}^{n}\frac{F_i^2}{\sigma_i^2} - \frac{\sqrt{\alpha}}{2}\sum_{i=1}^{n}\frac{F_i^{3/2}}{\sigma_i^2}bg_i - \frac{1}{2\sqrt{\alpha}}\sum_{i=1}^{n}\frac{m_i\sqrt{F_i}bg_i}{\sigma_i^2} + \frac{1}{2}\sum_{i=1}^{n}\frac{F_i bg_i^2}{\sigma_i^2} = 0 \quad (6)$$

The $\alpha$ parameter cannot be analytically evaluated at Eq. (6), since it cannot be obtained as a function of each one of the independent variables. For the required/obligatory numerical evaluation, *i*) an initial $\alpha$ value must be calculated, which can be obtained from Eq. (4). Later, it should be refined numerically; and *ii*) an arbitrary sequence of data $bg_i$ should be included, with the properties required at Eq. (5).

In order to obtain the numerical evaluation of the parameter $\alpha$ we apply to Eq. (6) a property of the spectroscopic measurements, where the standard deviation at each channel is $\sigma_i(m_i) = \sqrt{\alpha F_i}$. Then, we can write:

$$\sum_{i=1}^{n} m_i - \alpha\sum_{i=1}^{n} F_i - \frac{\sqrt{\alpha}}{2}\sum_{i=1}^{n}\sqrt{F_i}bg_i - \frac{1}{2\sqrt{\alpha}}\sum_{i=1}^{n}\frac{m_i bg_i}{\sqrt{F_i}} + \frac{1}{2}\sum_{i=1}^{n} bg_i^2 = 0 \quad (7)$$

The standard error propagation formula is required in order to evaluate the uncertainty of the parameter $\alpha$. This formula is expressed as:

---

[^T]: There are many denominations about the "ad hoc" modifications over the likelihood function. It is not clear in which category this proposition should be included.

$$\Delta^2 H(x_i) = \sum_{j=1}^{k} \left(\frac{\partial H(x_i)}{\partial x_i}\right)^2 (\Delta x_i)^2 \quad (8)$$

Being $\Delta H(x_i)$ the uncertainty of the function $H(x_i)$, and $\Delta x_i$ the uncertainty of every one of the $k$ independent input variables.

The Eq. (8) is implicitly applied over Eq. (6). Taking into account that $\Delta F_i = 0$ and $\Delta bg_i = 1$, the uncertainty of the parameter $\alpha$, is:

$$\Delta^2 \alpha = \frac{\left(\sum_{i=1}^{n}\left(\frac{F_i}{\sigma_i^2} - \frac{1}{2\sqrt{\alpha}}\frac{\sqrt{F_i}bg_i}{\sigma_i^2}\right)\Delta m_i\right)^2 + \left(\sum_{i=1}^{n}\left(\frac{\sqrt{\alpha}}{2}\frac{F_i^{3/2}}{\sigma_i^2} + \frac{1}{2\alpha}\frac{m_i\sqrt{F_i}}{\sigma_i^2} - \frac{F_i bg_i}{\sigma_i^2}\right)\Delta bg_i\right)^2}{\left(\sum_{i=1}^{n}\left(\frac{F_i^2}{\sigma_i^2} + \frac{1}{4\sqrt{\alpha}}\frac{F_i^{3/2}bg_i}{\sigma_i^2} + \frac{1}{4\alpha^{3/2}}\frac{bg_i m_i \sqrt{F_i}}{\sigma_i^2}\right)\right)^2} \quad (9)$$

The $\alpha$ parameter previously obtained at Eq. (7) is required for the evaluation of Eq. (9).

We applied this method over computer simulated spectra. The proposed method produces results with a notorious enhancement on the quality of the accuracy and precision of the results, compared with those provided by the least squares algorithms.

**APPLICATION EXAMPLES**

This method was applied to simulated data in order to evaluate its characteristics and the improvement obtained over the quality of the results.

We applied this method over an arbitrary sinusoidal function $F(x) = a \sin(x) + b$. In this particular case, the function is defined as:

$$F(x) = 27 \sin(x/32.3) + 17 \quad (10)$$

At each channel $i$, in a simulated spectroscopic acquisition the "measured" $mi$ signal is obtained from the $Fi$ function affected by fluctuations ruled by the Poisson´s statistics, as:

$$m_i = \alpha F_i + bg_i \sqrt{\alpha F_i} \quad (11)$$

At Eq.(11) a particular Gaussian random sequence $(bgi)1$ is required, having mean value zero and standard deviation $\Delta=1$.

The $Fi$ function and one of its possible "measured" $mi$ signals are shown in Fig. (1).

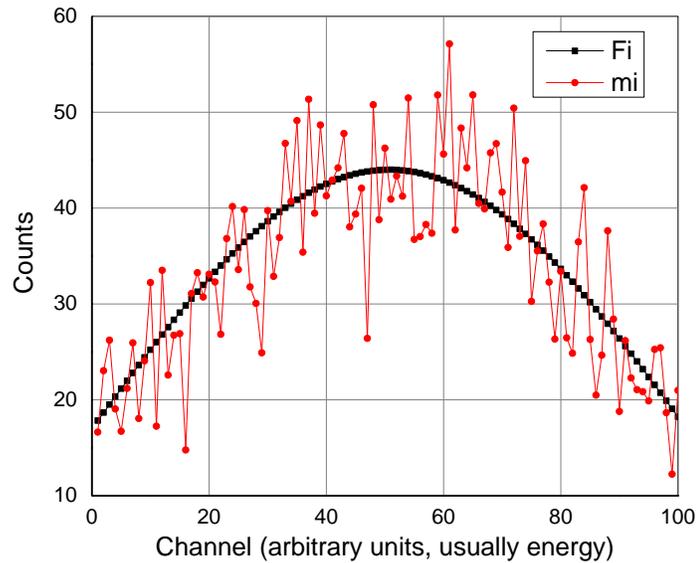

*Figure 1. A sinusoidal function ($F_i$, black points) and one of its simulated spectroscopic results ($m_i$, red points) obtained by an atomic spectroscopic technique. This simulated measurement is affected by fluctuations ruled by the Poisson´s statistics.*

The random sequence $(bg_i)1$ used in order to obtain the $m_i$ data is shown in the Fig. (2), where also a second sequence $(bg_i)2$ with the same properties is included. Both sets of data could be used in order to evaluate the Eq.(7).

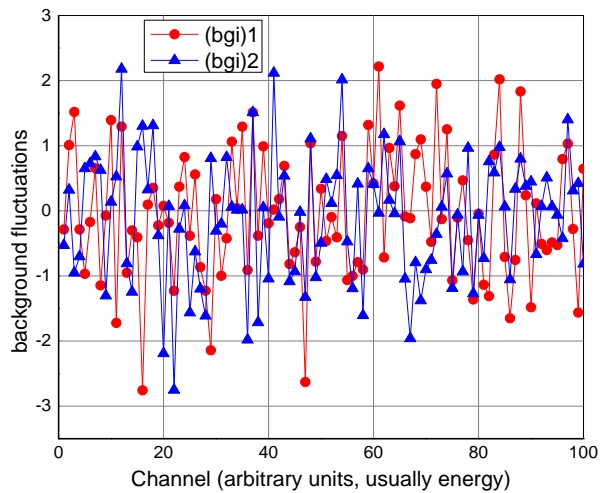

*Figure 2. Two sequences of arbitrary $bg$ data. The first set ($(bg)1$, red points) was used to construct the "measured" sequence $m_i$, shown in the Fig.(1). This sequence can be applied in Eq.(7). The second set ($(bg)2$, blue points) or any other Gaussian random sequence (with mean value zero and standard deviation Δ=1) also can be used in the Eq.(7) in order to evaluate the $α$ parameter.*

In the Fig. (1) the *mi* data is obtained from the *F* function when it is affected by a particular realization of the statistical fluctuations $(bgi)1$, so in this case the $\alpha$ parameter is strictly 1.

In order to check the proper development of the Eq.(7), we calculated the $\alpha$ parameter from the *mi* data shown in the Fig.(1). In this particular case the $(bgi)1$ set of data is used twice, first in order to construct the *mi* data from the *F* function, and second as the *bgi* set of data required in Eq.(7). An initial guess of the alpha value can be obtained applying the Eq. (4) to this $m_i$ data. In this case we get: $\alpha = 0.985 \pm 0.02$. Later we refine this value determining the most appropriate number which makes zero the second member of the Eq.(7). The numerical evaluation of the parameter $\alpha$ is shown in Fig. (3).

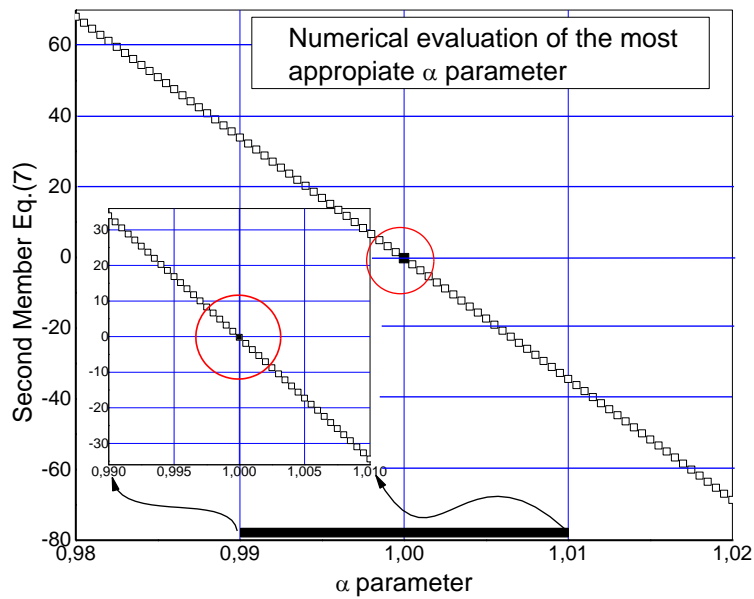

*Figure 3. Numerical evaluation of the most appropriate $\alpha$ value, which makes zero the second member of Eq. (7). In this ideal case, a) the exact $\alpha$ value is strictly 1 and b) the bg set of data is used twice, first to construct the mi data, and second as term required in Eq.(7). A zoom of the region of interest is included, where is observed that the alpha value is recovered with high accuracy.*

As can be seen from Fig.(3), the alpha value is recovered with high accuracy from the developed formulation. The very small difference obtained (~0.01%) can be attributed to numerical errors (round-off error). This is a characteristic result, with independency of the intensity of the $\alpha$ parameter, if the *bg* set of data is simultaneously used in order to produce the *mi* data and as a term in the Eq.(7).

But in a real measurement we don´t know which is the random noise that affects the pure signal. In this case, we should propose another set of *bg* data in order to evaluate Eq.(7), as is shown in the Fig.(2) (blue points). For the case being analyzed the alpha parameter obtained with this new *bg* data is $\alpha = 0.995$.

**Evaluation of the Accuracy and Precision**

For the evaluation of the accuracy and precision obtained with this method, in the Table (1) we show ten calculations of the $\alpha$ parameter from the $mi$ data shown in Fig.(1). In each case, we require a different $bg$ set of data in order to evaluate the Eq.(7). From the obtained $\alpha$ values we can estimate the quality of the results obtained with this method.

| #Evaluation | Real Value | | a) Eq.(4) | | b) Eq.(7) |
|---|---|---|---|---|---|
| | $\alpha$ | $\Delta\alpha$ | $\alpha$ | $\Delta\alpha$ | $\alpha$ |
| 1 | 1 | 0 | 0.985 | 0.02 | 0.9952 |
| 2 | | | | | 0.9871 |
| 3 | | | | | 0.9992 |
| 4 | | | | | 1.0157 |
| 5 | | | | | 1.0058 |
| 6 | | | | | 1.0215 |
| 7 | | | | | 1.0105 |
| 8 | | | | | 0.9895 |
| 9 | | | | | 1.0017 |
| 10 | | | | | 1.0096 |

*Table 1. Ten evaluations of the $\alpha$ parameter from the same $mi$ data. Case a) Evaluation from a previously developed least squares method, Eq.(4). b) Results obtained from the proposed method, Eq.(7); in each case a different realization of random noise is required as the $bg$ term.*

The results at Table 1 are also shown at Fig.(3), where is observed that usually this method produces better results than the least squares approach.

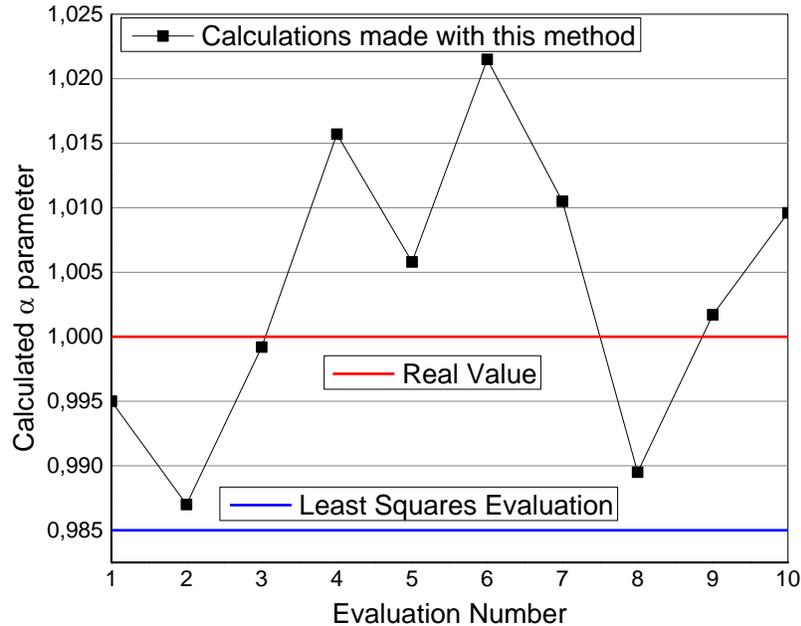

*Figure 3. Evaluations of the α parameter from the same $mi$ data. a) Real value of the α parameter (red line); b) Evaluation from a least squares approach (blue line), and c) Ten results obtained from the proposed method (black lines), for each case a different realization of random noise is required as the $bg$ term, in order to evaluate the Eq.(7).*

From the same $mi$ data we can obtain a large number of evaluations. For each evaluation, a new sequence of $bg$ data is required. In this case the averaged ten results in Fig.(3) produce the value, $\alpha = 1.0035 \pm 0.011$.

We observe that the quality of the accuracy and precision in the results obtained with this method, are improved one order of magnitude, compared with those produced by the least squares algorithms. Since more evaluations can be simply obtained, the quality of the results can be improved easily (in this ideal system).

**CONCLUSIONS**

In data spectral analysis, the application of the Maximum Likelihood formulation is almost perfectly suited, since it is constructed based on a deep agreement between the underlying physical phenomena studied and the applied mathematical theory. In this sense, the values obtained from the classical least squares approach should be the best, and no method of improvement of the results could be proposed. Moreover, the properties of the obtained results were largely studied: *i)* The results from the least squares formulation are unbiased (in the limit of infinite measurements) and *ii)* they have minimum variance among all unbiased linear estimators. This means that the estimates "get us as close to the true unknown parameter values as we can get". For these reasons, an improvement on the quality of the results over those obtained by least squares approaches seems to be unrealistic.

However, in another work we developed a new smoothing method [6] which was applied to simulated spectroscopic data, producing results with better accuracy than those obtained from the least squares approach.

In this paper, we propose a modification in the construction of the Likelihood function, which leads to a remarkable improvement on the quality of the results. This modification was made taking into account the underlying statistics of the phenomena studied, by the inspection of the residues of the fitting, which should behave with specific statistical properties. This new formulation was analytically developed, but the calculated parameter should be evaluated numerically, since it cannot be obtained as a function of each one of the independent variables. For the required numerical evaluation, we require: *i*) the sought signal $F$, to be evaluated from the acquired spectrum, should be accurately known, *ii*) along with the acquired spectrum, we should process many sets of external data with specific properties. This arbitrary term is a random sequence of data which is uncorrelated with the acquired signal. It should be ruled by a Gaussian distribution, having mean value zero and standard deviation $\Delta=1$.

As was shown, the results obtained from Maximum Likelihood can be improved, but a greater number of data should be handled. In this method, if we made $n$ evaluations, the total number of data processed is the $n$ times greater than the number of the acquired signal – this external data is artificially inserted in the numeric procedure. But, if we only process the acquired data, the quality of the results from least squares approach cannot be enhanced.

The developed statistical method was evaluated using computer simulated spectra. The numerical estimations of the calculated parameter applying this method, indicate an improvement over accuracy and precision, one order of magnitude better than those produced by the common least squares approaches (see Table 1). Since more evaluations can be simply obtained, the quality of the results can be improved easily (in this ideal system).

In spectral analysis, we still have to evaluate the improvement produced by this method in conjunction with a new smoothing procedure [6], over Detection and Quantitation Limits, when they are applied over real experimental results. These kind of evaluations are formally required in all spectroscopic techniques, in order to establish their metrological traceability, defining carefully their operational procedures. For instance, in TXRF spectroscopy there is a list of reports which evaluate its characteristics.[15,16,17,18] We consider that this method will improve the quality of the results in many spectroscopic techniques and related systems.